\pdfoutput=1

\documentclass[ 10pt, english, a4paper, onecolumn, oneside ]{article}

\usepackage{etoolbox}
\DeclareTextFontCommand \textti   {\fontfamily{cmdh}\selectfont}
\DeclareTextFontCommand \textsfbf {\sffamily\bfseries}

\newlength{\ptsize}
\usepackage[sf,bf,compact,outermarks,pagestyles,newlinetospace]{titlesec}

\titlelabel{\llap{\thetitle\hspace*{\ptsize}}}

\RequirePackage{enumerate}

\usepackage{localconfig}

\newtagform{sftag}[\textsf]{(}{)}
\newtagform{triangletag}[\textsf]{\raisebox{1pt}{\ensuremath{\displaystyle\triangleright}}\space}{}
\usetagform{triangletag}

\hypersetup{
  hidelinks,
  breaklinks=true,
  pdfborder={0 0 0},
  pdftitle={Average Consensus: A Little Learning Goes A Long Way},
  pdfauthor={Bernadette Charron-Bost, Patrick Lambein-Monette},
  pdfkeywords={average consensus, dynamic networks, distributed algorithms, iterated averaging, EqualNeighbor update rule, Metropolis update rule},
  pdfcreator={LaTeX with lipics-v2019.cls and hyperref.sty},
  pdfsubject={Average consensus in dynamic networks},
}

\begin{document}

\begin{center}
\begin{minipage}{0.618\textwidth}
\LARGE\textti{Average Consensus:\\*
\hspace*{\fill}A Little Learning \hspace*{\fill}\\*
\hspace*{\fill}Goes A Long Way}
\end{minipage}
\end{center}

% Authors section
\vspace*{\parskip}
\begin{center}
\renewcommand{\thefootnote}{\fnsymbol{footnote}}
\begin{minipage}[t]{0.49\textwidth}
\noindent\textsf{Bernadette Charron-Bost}%
\footnotemark[1]\footnotemark[2]\\*
{\small\email{charron@lix.polytechnique.fr}}
\end{minipage}
\begin{minipage}[t]{0.49\textwidth}
\noindent\textsf{Patrick Lambein-Monette}\footnotemark[2]\\*%
{\small\email{patrick@lix.polytechnique.fr}\\*
\makebox[0pt][r]{\orcidlogo[7]\space}\orcidurl{0000-0002-9401-8564}}
\end{minipage}
\footnotetext[1]{CNRS}
\footnotetext[2]{LIX, École polytechnique, Institut Polytechnique de Paris}
\renewcommand{\thefootnote}{\arabic{footnote}}
\setcounter{footnote}{0}
\end{center}

\noindent
\makebox[0pt][r]{\textsf{\textbf{Abstract}}\hspace*{\ptsize}}%
  When networked systems of autonomous agents
    carry out complex tasks,
    the control and coordination sought after
    generally depend on a few fundamental control primitives.
  % more detailed background
  Chief among these primitives is \reintro{consensus},
    where agents are to converge to a common estimate
    within the range of initial values, % private to each agent,
    which becomes \reintro{average consensus}
    when the joint limit should be %is guaranteed to be
    the average of the initial values.
  To provide reliable services that are easy to deploy,
    these primitives should operate
    even when the network is subject
    to frequent and unpredictable changes.
  Moreover, they should mobilize few computational resources
    so that low powered, deterministic, and anonymous agents
    can partake in the network.

% general problem
  In this stringent adversarial context,
    we investigate the distributed implementation of these primitives
    over networks with bidirectional,
    but potentially short-lived, communication links.
  % summary of the main result (“here we show”)
  Inspired by the classic EqualNeighbor and Metropolis agreement rules
    for multi-agent systems,
    we design distributed algorithms
    for consensus and average consensus,
    which we show to operate in polynomial time
    in a synchronous temporal model.
  These algorithms are fully distributed,
    requiring neither symmetry-breaking devices
    such as unique identifiers,
    nor global control or knowledge of the network.
  Our strategy consists
    in making agents learn simple
    structural parameters of the network
    -- namely, their largest degrees --
    which constitutes enough information
    to build simple update rules,
    implementable locally with little computational and memory overhead.

\section{Introduction}

\subsection{Controlling networked systems}

\textsc{From} an initial state of discord,
  a group is said to reach agreement
  when everyone eventually adopts the same opinion.
This problem is central in multi-agent
  systems whether they be natural or artificial.
Examples of the latter kind include
  flocks of autonomous vehicles or drones,
  distributed ledgers,
  smart power grids,
  decentralized contact-tracing apps,
  and sensors for the remote monitoring
  of environmental or industrial parameters,
  to name some vibrant areas
  in which applications are being developed.

We consider a networked system of $n$ \emph{agents}
  -- a generic term we use to denote
  the autonomous nodes of the network
  regardless of their nature:
  robots, cellular phones, sensors\ldots --
  denoted by the integer labels $ 1, \ldots, n $.
Each agent~$i$ starts with a value $\mu_i$,
  which we call its \reintro{input}
  and is taken arbitrarily from the domain of the problem,
  assumed here to be the set of real numbers $\bbR$.
The input represents the private observation made by the agent
  of some aspect of the environment
  that is relevant to the task at hand:
  for example a temperature reading,
  or the direction, speed, or location of the agent if it is mobile.
We focus on two flavors of a control primitive
  where the entire network strives to reach a common value
  that is compatible with the input values~$\mu_1, \ldots, \mu_n$
  in a sense we now make precise.

The first one is \reintro{consensus}:
  each agent~$i$ maintains over time
  its own estimate~$x_i(t)$ of the objective,
  and the estimates should asymptotically converge
  to some common limit $\omega$
  within the range of the input values
  -- we say that the estimates
  achieve \defterm{asymptotic consensus} over~$ \omega $.
Depending on the system being modeled,
  the estimate may be an internal variable
  storing an intermediate result
  of some long computation,
  but it may also directly measure
  some relevant parameter controlled by the agent,
  such as its speed or position.

Among many practical applications,
  a consensus primitive can serve
  to coordinate mobile agents:
  have them regroup,
  adopt a common heading,
  or control their relative positions while moving together
  -- problems formally known as \emph{rendez-vous}%
  ~\cite{alpern1995sjco, suzuki1999sjc},
  \emph{flocking}~\cite{vicsek1995prl,
  hendrickx2006p1ismtns, cucker2007itac},
  and \emph{formation control}~\cite{balch1998itra, ren2007icta}.
Examples not grounded in mobile agents include
  implementing a distributed clock~\cite{lamport1985ja,
  welch1988iac,
  qunli2006itc,
  simeone2007jwcn},
  or indeed in natural systems
  to get cardiac pacemaker cells working in concert
  to produce heartbeats~\cite{mirollo1990sjam, daliot2003ss}
  or fireflies to flash in unison~\cite{buck1938tqrob, mirollo1990sjam}.

The second primitive, \reintro{average consensus},
  adds the constraint that the common limit $\omega$
  should be the arithmetic mean of the input values
  $\avgvec{\vec{\mu}} = \frac{1}{n}(\mu_1 + \cdots + \mu_n)$.
This is sometimes required for specific applications
  such as sensor fusion~\cite{xiao2005i2fisipsn2},
  load balancing~\cite{cybenko1989jopadc, dinitz2017ii2-iccc},
  or distributed optimization and machine learning~\cite{%
    olshevsky2017sjco, nedic2018pi}.

We look at these primitives
  in the adversarial context of \emph{dynamic networks},
  where the communication links joining the agents evolve over time.
Indeed, agreement primitives are often required
  in settings that are inherently dynamic
  and unadequately described by static networks:
  we cite for example mobile ad hoc networks,
  where the links change
  as the agents move in space due to external factors;
  autonomous vehicular networks,
  where again change is caused by mobility,
  but which results this time from our control;
  or peer-to-peer networks,
  which continuously reconfigure
  as agents join and leave.

A standard approach to consensus
  has agents regularly adjust their estimates
  as a convex combination of those of neighboring agents,
  defined by a \emph{convex update rule}.
We adopt a temporal model of \emph{synchronized rounds},
  and this can be rephrased,
  in each round~$t$ and for each agent~$i$,
  as an update of the general form
  $x_i (t) = \sum_{j \in \Ns{i}{t}} a_{ij} (t) x_j (t - 1)$,
  where the weights $a_{ij} (t)$
  are taken to form a convex combination,
  and the sum is over
  the incoming neighbors of agent~$i$ in
  the communication graph in round~$t$.

\subsection{Contribution}

We study such updates from a computational angle:
  what sort of weights $a_{ij}(t)$
  can be computed locally by the agents
  and produce good behavior?
For a simple example,
  given enough connectivity,
  average consensus comes easily
  if we make each agent~$i$ pick
  $a_{ij}(t) = \frac{1}{n}$
  for its proper neighbors $j$,
  and picks $a_{ii}(t) = \frac{\degree{i}{t}-1}{n}$ for itself,
  where the degree $\degree{i}{t}$ is the count
  of its neighbors, itself included,
  in the communication graph in round~$t$.
However simple this scheme is to describe,
  picking these weights in a distributed manner
  requires agents to know~$n$,
  but evaluating the network size
  in a distributed manner
  is itself an entirely non-trivial feat.
We will argue that the Metropolis rule --
  where the weights for neighbors neighboring agents $ i $ and $ j $
  are given by
  $ a_{ij}(t) = \frac{1}{\max(\degree{i}{t},\degree{j}{t})} $ --
  breaks down over dynamic networks
  because of similar issues.
On the other hand,
  implementing the EqualNeighbor weights
  $a_{ij}(t) = \frac{1}{\degree{i}{t}}$
  poses no problem,
  but convergence over dynamic networks
  can be dramatically slower than in the static case.

Our main contribution
  consists in a pair of algorithms
  for consensus and average consensus,
  the MaxWeight and MaxMetropolis algorithms,
  which operate over dynamic networks
  as long as the communication links are bidirectional,
  but without relying on some central control (e.g., identifiers or leaders)
  or global knowledge (e.g., a bound on the network size).
The corresponding update rules depend
  on structural parameters of the dynamic network,
  learned along the way by each agent
  -- namely, its largest degree in the communication topology.
Our strategy induces a moderate delay
  in the temporal complexity
  when compared to the static case
  for the EqualNeighbor and Metropolis rules:
  these rules have respective temporal bounds in
  $\bigo{\cramped{n^3 \log n / \varepsilon}}$
  and $\bigo{\cramped{n^2 \log n / \varepsilon}}$,
  while both our algorithms admit a bound in
  $\bigo{\cramped{n^4 \log n / \varepsilon}}$.

We establish these complexity bounds
  with geometric arguments
  relying on spectral graph theory,
  but each update rule is subject
  to its own specific challenges.
The MaxMetropolis update is not convex
  throughout the entire execution,
  causing the system to away from consensus,
  and inducing a delay that we have control.
The MaxWeight update faces a subtler issue:
  different geometries best quantify its progress at different times,
  but switching geometries wipes off much of said progress,
  and we have to show that the succession of those switches
  cannot stall convergence too much.

\subsection{Related works}

Convex update rules are closely related
  to random walks over finite graphs,
  whose origins go back to the development
  of the field of probability theory itself.
For a formulation and purpose that resemble ours,
  we mention early works in opinion dynamics by
  DeGroot~\cite{degroot1974jasa},
  Chatterjee~\cite{chatterjee1975p4sisiwp1},
  and Chatterjee and Seneta~\cite{chatterjee1977jap},
  followed later by works on distributed control
  by Tsitsiklis~\cite{tsitsiklis1984a}
  and Tsitsiklis et al.~\cite{tsitsiklis1986itac}.
Modern interest in the matter,
  and especially in the EqualNeighbor rule,
  can be traced back to the work on the simulation of flocking
  done by Reynolds~\cite{reynolds1987cg}
  and then by Vicsek et al.~\cite{vicsek1995prl},
  followed by analytical analyses by
  Jadbabaie et al.~\cite{jadbabaie2003itac}
  and by Olfati-Saber and Murray~\cite{olfati-saber2004itac},
  after which the sustained attention
  that the topic has received
  precludes anything resembling a complete bibliographical review;
  see~\cite{
    xiao2004s&cl,
    moreau2005itac,
    blondel2005p4icdc,
    hendrickx2006p1ismtns,
    cao2008sjco,
    olshevsky2011sr,
    nedic2009itac,
    chazelle2011sjco,
    charron-bost2015alpi2,
    nedic2017itac}
  for a few more milestones.

Update rules for consensus and average consensus
  generally require the weights $ a_{ij}(t) $ to be non-negative
  so that they form a convex combination.
A notable exception is found in~\cite{xiao2004s&cl},
  where Xiao et al. look for the \emph{fixed} weights $ a_{ij} $
  that optimize the speed of convergence over a given fixed graph,
  and find that the weights may be negative.
The MaxMetropolis algorithm
  is itself able to solve average consensus over dynamic networks
  precisely because its update weights can sometimes be negative.
When compared with our approach,
  the important difference is that
  we consider dynamic graphs
  and focus on distributed implementation of
  the update rules,
  while the weights obtained in~\cite{xiao2004s&cl}
  are given by a centralized optimization problem
  and are incompatbile with a distributed approach.

A number of strategies aim at
  speeding up convex update rules over static networks
  by having the agents learn what amounts to
  spectral elements of the graph laplacian~\cite{charalambous2016itcns},
  which can dramatically improve convergence~\cite{sundaram20072acc}.
Like our own algorithms,
  these represent distributed methods
  by which the agents learn structural properties
  of the communication graph.
However, these methods need agents
  to be issued unique identifiers,
  and they are gluttonous in memory
  and heavy in computation,
  with agents computing and memorizing, in each rounds,
  kernels of Hankel matrices of dimension
  $\mathrm{\Theta}(n) \times \mathrm{\Theta}(n)$.
In contrast, our method works with anonymous agents,
  uses $\lceil \log n \rceil$ bits of memory and bandwidth,
  and has a trivial computational overhead.

Average consensus update rules for dynamic networks
  were notably studied by Olshevsky and Tsitsiklis~\cite{olshevsky2011sr}
  -- which defined the convergence time
  and convergence rate of an algorithm --
  a work later expanded
  together with Nedić and Ozdaglar in~\cite{nedic2009itac}.
These analytical approaches focus on aspects
  such as the temporal complexity
  and tolerance to quantization,
  whereas we address issues of a distributed nature,
  in particular the implementation of rules
  by distributed algorithms.

We now proceed along the following structure.
In \Cref{sec:preliminaries},
  we recall some mathematical terminology and known results,
  and detail our computational model.
\Cref{sec:averaging} presents update rules,
  in particular the EqualNeighbor rule for consensus,
  and the Metropolis rule for average consensus,
  and discusses their limitations.
We give the MaxWeight consensus algorithm in \Cref{sec:mw},
  detailing the underlying learning strategy,
  which we then apply in \Cref{sec:mm}
  to the average consensus problem
  with the MaxMetropolis algorithm.
We conclude in with a discussion in \Cref{sec:discussion}.

\section{Preliminaries}
\label{sec:preliminaries}

\subsection{Mathematical toolbox}\label{subsec:mathtools}

Let us fix some notation.
If $ k $ is a positive integer,
  we denote by $ [ k ] $ the set
  $\set{1, \ldots , k}$.
If $ S $ is any non-empty finite set of real numbers,
  we denote its \defterm{diameter} by
  $ \diam S \defeq \max S - \min S $.

A \defterm{directed graph} (or simply \reintro{graph})
  ~$ G = (V , E) $
  with vertices in~$ V $ and edges in~$ E \subseteq V \times V $
  is called \defterm{reflexive}
  when $ (i , i) \in E $ for all $ i \in V $;
  $ G $ is \defterm{bidirectional} when
  $ ( i , j ) \in E \iff ( j , i ) \in E $
  for all $ i , j \in V $;
  and $ G $ is \defterm{strongly connected}
  when directed paths join any pair  of vertices.

  % TODO: remove "order"
All graphs that we consider here
  will be reflexive, bidirectional,
  and strongly connected graphs
  of the form $ G = ([n] , E) $.
In such a graph,
  the vertices linked to a vertex~$i$
  form its \defterm{neighborhood}
  $\Ns{i}{G} \defeq \set{j \in [n] \where (j,i) \in E}$,
  and the count of its neighbors
  is its \defterm{degree}
  $\degree{i}{G} \defeq \abs{\Ns{i}{G}}$.
By definition, the degree is at most~$ n $,
  and in a reflexive graph it is at least~$ 1 $.

Matrices and vectors will be consistently denoted
  in bold italic style:
  upper case for matrices ($\mat{A}, \mat{B}$)
  and lower case for vectors ($\vec{u}, \vec{v}$),
  with their individual entries
  in regular italic style:
  $A_{ij}, B_{jk}, u_i, v_j$.
We use the shorthand $ \dvec{v} $
  to denote the infinite vector sequence
  $ \vec{v}(0) , \vec{v}(1) , \cdots $.

The graph~$ G_{\mkern-2mu\mat{A}} = ( [n] , E ) $ \defterm{associated}
  to a matrix $ \mat{A} \in \bbR^{ n \times n } $
  is defined by
  $ ( j , i ) \in E \iff A_{ij} \neq 0 $
  for all $ i , j \in [n] $.
The matrix~$ \mat{A} $ is said to be \defterm{irreducible}
  when~$ G_{\mkern-2mu\mat{A}} $ is strongly connected.

Given a vector $ \vec{v} \in  \bbR^{n}$,
  we write $ \diam \vec{v} $
  to mean the diameter of the \emph{set}~%
  $ \set{ v_1, \cdots, v_n } $ of its entries.
The diameter constitutes a seminorm over $\bbR^n$;
  in particular, we  have $ \diam \vec{v} \geqslant 0 $.
We call \defterm{consensus vectors} those of null diameter;
  they form exactly the linear span of the constant vector
  $\vec{1} \defeq (1,1,\cdots,1)^\transp$.
The \defterm{identity matrix}~$\idmat$
  is the diagonal matrix
  $ \operatorname{diag} ( \vec{1} ) $.

A matrix or a vector with non-negative (resp. positive) entries
  is itself called \defterm{non-negative} (resp. positive).
A non-negative vector is called \defterm{stochastic}
  if its entries sum to $1$.
A non-negative matrix $\mat{A}$ is \reintro{stochastic}
  if \emph{each of its rows} sums to $1$ --
  equivalently, if $\mat{A} \vec{1} = \vec{1}$.
If moreover its transpose $ \mat{A}^{\transp} $ is stochastic,
  $ \mat{A} $ is \defterm{doubly stochastic},
  and it is the case in particular
  when $ \mat{A} $ is \defterm{symmetric}.

We denote the mean value of a vector $ \vec{v} \in \bbR^n $
  by $\avgvec{\vec{v}} \defeq \frac{1}{n} \sum_{i} v_i$.
Doubly stochastic matrices play a central role
  in the study of average consensus,
  as multiplying any vector $ \vec{v} $
  by a doubly stochastic matrix $ \mat{A} $
  preserves its average -- that is,
  $ \avgvec{\mat{A} \vec{v}} = \avgvec{\vec{v}} $.

For any matrix $ \mat{A} \in \bbR^{ n \times n } $,
  we can arrange its $ n $ eigenvalues
  $ \lambda_1, \ldots, \lambda_n $,
  counted with their algebraic multiplicities,
  in decreasing order of magnitude:
  $
    \abs{\lambda_1} \geqslant
    \abs{\lambda_2} \geqslant
    \cdots          \geqslant
    \abs{\lambda_n}
  $.
Under this convention,
  the \defterm{specral radius} of the matrix $ \mat{A} $
  is the quantity
  $ \sprad{\mat{A}} \defeq \abs{\lambda_1} $,
  and its \defterm{spectral gap} is the quantity
  $ \spgap{\mat{A}} \defeq \abs{\lambda_1} - \abs{\lambda_2} $.
In particular, a stochastic matrix has a spectral radius of $ 1 $,
  which is itself an eigenvalue
  that has $ \vec{1} $ for eigenvector.

\subsection{Computing model}

We consider a networked system of $ n $ agents, denoted $1, 2, \ldots, n$.
Computation proceeds in \emph{synchronized rounds}
  that are communication closed,
  in the sense that no agent receives messages in round $t$
  that are sent in a different round.
In round $t$ ($t = 1, 2 , \ldots $),
  each agent~$ i $ successively
  \begin{enumerate*}[
    label     = \textbf{\sffamily\alph*)},
    itemjoin  = {{; }},
    itemjoin* = {{; and }}]
  \item broadcasts a \emph{single} message $ m_i(t) $
    determined by its state at the beginning of round~$ t $%
    \label{enum:send}%
  \item receives \emph{some} of the messages $ m_1(t), \ldots, m_n(t) $%
    \label{enum:adversary}%
  \item undergoes an internal transition to a new state%
    \label{enum:transition}%
  \item produces a \emph{round output} $ x_i(t) \in \bbR $
    and proceeds to round~$t+1$.%
    \label{enum:output}%
  \end{enumerate*}
Communications that occur in round~$t$
  are modeled by a directed graph~$\bbG(t) \defeq ([n],  E(t))$,
  called the round~$t$ \defterm{communication graph},
  which may change from one round to the next.
We assume each communication graph $ \bbG(t) $ to be reflexive,
  as agents always know their own messages.

Messages to be sent in step~\ref{enum:send}
  and state transitions in step~\ref{enum:transition} are determined
  by a \emph{sending} and a \emph{transition} functions,
  which together define the \emph{local algorithm}
  for agent~$i$.
Collected together, the local algorithms
  of all agents in the system
  constitute a \defterm{distributed algorithm}.

An \defterm{execution} of a distributed algorithm
  is a sequence of rounds, as defined above,
  with each agent running the corresponding local algorithm.
We assume that all agents start simultaneously in round~$ 1 $,
	since the algorithms under our consideration
  are robust to asynchronous starts,
  retaining the same time complexity
  as when the agents start simultaneously.
Indeed, asynchronous starts only induce
  an initial transient period
  during which the network is disconnected,
  which cannot affect the convergence and complexity
	results of algorithms driven by convex update rules.

The entire sequence $ \dvec{x} $ in any execution of an algorithm
  is entirely determined by the input vector $ \vec{\mu} $
  and the patterns of communications in each round~$ t $, i.e.,
  the sequence of communication graphs
  $ \bbG \defeq \cramped{{(\bbG(t))}_{t \geqslant 1}} $,
  called the \defterm{dynamic communication graph} of the execution,
  and so we write $ \dvec{x} = \dvec{x}( \bbG , \vec{\mu} ) $.
When the dynamic graph~$ \bbG $ is understood,
  we let $ \Ns{i}{t} $ and $ \degree{i}{t} $
  respectively stand for
  $ \Ns{i}{\bbG(t)} $ and $ \degree{i}{\bbG(t)} $.
As no confusion can arise,
  we will sometimes identify an agent
  with its corresponding vertex in the communication graph,
  and speak of the degree or neighborhood of an \emph{agent}
  in a round of an execution.

We call a \defterm{network class}
  a set of dynamic graphs;
  given a class $ \grcl{C} $,
  we denote by $ \grcl{C}_{\vert n} $
  the sub-class $ \set{ \bbG \in \grcl{C} \where \abs{\bbG} = n } $.
In the rest of the paper,
  we will center our investigation on the class $ \grcl{G} $
  of dynamic graphs of the following sort.

\begin{assumption}
  \label{assumption:graphs}
  In each round~$t \in \bbNz$,
    the communication graph~$\bbG(t)$
    is reflexive, bidirectional, and strongly connected.
\end{assumption}

\section{Update rules for asymptotic consensus}
\label{sec:averaging}

Our approach distinguishes local algorithms,
  as defined above,
  from the \emph{update rules} that they implement:
  the latter are recurring formulas
  describing how the estimates $ x_i(t) $ change over time,
  while the former specify
  the distributed implementation of such rules
  with local algorithms
  -- that is, with each agent only using the information
  that is locally available.
% This discrepancy has been pointed out in~\cite{CBM19}
%   for the \emph{stabilizing consensus} problem,
%   where they consider \emph{MinMax} update rules,
%   and show that \emph{MinMax} algorithms
%   can implement such rules in a distributed manner
%   by having each agent maintain an auxiliary array of size~$ n $.
When we consider the EqualNeighbor rule,
  which involves purely local parameters
  and directly corresponds to a local algorithm,
  this distinction may appear exacting.
However, it becomes crucial
  when we consider the Metropolis rule,
  which is easily defined,
  but challenging to implement
  because of its dependence on information
  in the neighborhood at distance~$ 2 $ of each agent.
Collecting this information in a distributed manner
  can only be done over specific graph classes,
  in a manner that does not generalize
  to the entire class $ \grcl{G} $.

\subsection{Update rules and convergence}

We will focus on distributed algorithms which realize,
  in their executions,
  affine update rules of the general form
  \begin{equation}
    \label{eq:update_rule}
    x_{i}(t) = \sum_{\mathclap{ j \in \Ns{i}{t} }}
                  a_{ij}(t) \mkern2mu x_j(t-1)
    \enspace,
    \qquad\qquad \text{with} \quad
    \sum_{\mathclap{j \in \Ns{i}{t}}} a_{ij}(t) = 1
    \enspace,
  \end{equation}
  where the time-varying affine weights $ a_{ij}(t) $
  may depend on the dynamic graph~$ \bbG $
  and the input vector $ \vec{\mu} $.
We then say that the algorithm \defterm{implements} the rule,
  but we insist again that
  \emph{an algorithm is not the same thing as a rule}.

Under our assumptions,
  the convergence of update rules,
  including all those we consider here,
  is ensured by the following classic result,
  found in the literature under various forms%
  ~\cite{chatterjee1977jap,
    tsitsiklis1984a,
    jadbabaie2003itac,
  moreau2005itac}.

\begin{proposition}
  \label{prop:convergence}
  Let $ \dvec{x} $ be a vector sequence
    satisfying the recurrence relation \cref{eq:update_rule}
    for weights $ a_{ij}(t) \geqslant \alpha > 0 $ --
    that is, the weights are positive
    and uniformly bounded away from $ 0 $.
  Under \Cref{assumption:graphs},
    the vectors $ \vec{x}(t) $ converge
    to a consensus vector.
\end{proposition}

We speak of \defterm{uniform convexity}
  when such a parameter $ \alpha > 0 $ exists,
  and we note that in that case \Cref{assumption:graphs}
  is much stronger than is necessary to achieve asymptotic consensus:
  as shown in~\cite{moreau2005itac},
  it suffices that the network
  never become permanently split.

We measure progress towards consensus
  with the \defterm{convergence time}:
  for a single sequence $ \dvec{z} $,
  it is given for any $ \varepsilon > 0 $ by
  $ \cramped{ \cvtime^{\dvec{z}}(\varepsilon) }
    \defeq \inf \set{
      t \in \bbN \where \forall \tau \geqslant t :
      \diam \vec{z}(\tau) \leqslant \varepsilon }
  $.
For an update rule or an algorithm,
  we measure in fact the \emph{worst-case} convergence time
  over a class~$ \grcl{C} $,
  defined for a system of size $ n $ by:
  \begin{equation}
    \cvtime[n]^{\grcl{C}} ( \varepsilon ) \defeq
    \mkern6mu\sup_{
      \crampedclap{\bbG \in \grcl{C}_{\vert n} ,\:  \vec{\mu} \in \bbR^n}
    }\mkern6mu
    \cramped{
      \cvtime^{\dvec{x}(\bbG,\vec{\mu})}
      ( \varepsilon \cdot \diam \vec{\mu} )
    } \enspace.
  \end{equation}
By default, we will consider the class $ \grcl{G} $,
  dropping the superscript and writing
  $ \cramped{\cvtime[n]^{\vphantom{\grcl{G}}}}(\varepsilon) $ for
  $ \cramped{\cvtime[n]^{\grcl{G}}}(\varepsilon) $.

\subsection{Example rules}

\subsubsection{EqualNeighbor}

The prototypical example of an update rule
  is the \defterm{EqualNeighbor} rule,
  where the next estimate of an agent
  is the unweighted average of those of its incoming neighbors:
  \begin{equation} \label{eq:en}
    x_{i}(t)
    = \frac{1}{\degree{i}{t}}
    \sum_{j \in \Ns{i}{t}} x_{j}(t-1)
    \enspace.
  \end{equation}
This rule directly translates into an implementing local algorithm:
  an agent broadcasts its latest estimate in every round,
  and picks for new estimate the average of its incoming values.
Since an agent's degree is at most $ n $,
  the EqualNeighbor rule is uniformly convex
  with parameter $ \alpha = 1/n $,
  and under \Cref{assumption:graphs}
  this algorithm solves consensus
  over the class $ \grcl{G} $.

However, the convergence \emph{time} is poor
  over the class $ \grcl{G} $,
  where~\cite[Proposition 12]{olshevsky2013itac}
  gives a \emph{lower} bound of
  $ \cvtime[n] (\varepsilon) \geqslant
    2^{\bigomega{n}} \log 1/\varepsilon $.
The convergence time is improved
  when the network displays additional structure:
  over the sub-class of $ \grcl{G} $ of \emph{fixed} graphs,
  Olshevsky and Tsitsiklis~\cite[Corollary 5.2]{olshevsky2011sr}
  use a result from~Landau and Odlyzko
  to show a tight bound in
  $ \cramped{\bigo{ n^3 \log n / \varepsilon }} $.
This bound extends to the sub-class $ \grcl{G} $
  of dynamic graphs for which each vertex
  has a fixed \emph{degree}~\cite[Theorem 1.6]{chazelle2011sjco}.

Local update rules such as \cref{eq:update_rule}
  admit an equivalent matricial form.
For the EqualNeighbor rule, as an example,
  \cref{eq:en} is equivalent to the global rule
  $ \vec{x}(t) = \mat{W} (\bbG(t)) \vec{x}(t-1) $,
  with the EqualNeighbor matrix $ \mat{W}( G ) $
  given for any graph $ G $ by
  \begin{equation}
  [\mat{W} ( G )]_{ij} =
    \begin{cases}
      1/\degree{i}{G} & j \in \Ns{i}{G}     \\
      0               & j \notin \Ns{i}{G} \enspace.
    \end{cases}
  \end{equation}
We note that this matrix is stochastic for any graph $ G $,
  and has $ G $ for associated graph.

\subsubsection{Metropolis}

Let us call the \defterm{Metropolis-Hastings symmetrization}
  the transform $ \mat{A} \mapsto \mathrm{M}(\mat{A}) $
  which, to any square matrix $ \mat{A} $,
  associates the matrix given by
  \begin{equation}
    \label{eq:metropolis}
    [\mathrm{M}(\mat{A})]_{ij} =
    \begin{cases}
      \min(A_{ij},A_{ji}) & j \neq i \\
      1 - \sum_{k \neq i} \min(A_{ik},A_{ki}) & j = i \enspace.
    \end{cases}
  \end{equation}

By construction,
  the matrix $ \mathrm{M}(\mat{A}) $ is symmetric
  and leaves consensus vectors invariant.
Outside the main diagonal,
  we have $ [\mathrm{M}(\mat{A})]_{ij} \leqslant A_{ij} $,
  and thus on the main diagonal we have
  $ [\mathrm{M}(\mat{A})]_{ii} \geqslant A_{ii} $;
  the matrix $ \mathrm{M}(\mat{A}) $
  is therefore doubly stochastic
  whenever the matrix $ \mat{A} $ is stochastic.

Xiao et al.~\cite{xiao2005i2fisipsn2}
  propose to approach the average consensus problem
  with the \defterm{Metropolis} update rule:
  \begin{equation}\label{eq:metropolis-local}
    x_i(t) = x_i(t-1)
    + \sum_{j \in \Ns{i}{t}}
    \frac{x_j(t-1) - x_i(t-1)}{\max(\degree{i}{t-1}, \degree{j}{t-1})}
    \enspace;
  \end{equation}
  when viewed at the scale of the system,
  this rule corresponds to symmetrizing the EqualNeighbor rule
  with \cref{eq:metropolis} round-wise, hence its name.
\Cref{prop:convergence} ensures
  asymptotic consensus of the Metropolis rule
  over the entire class $ \grcl{G} $
  as it did for the EqualNeighbor rule,
  and since the EqualNeighbor matrices are stochastic,
  the Metropolis matrices are doubly stochastic,
  and the Metropolis rule results
  in fact in \emph{average} consensus,
  with a convergence time in
  $ \bigo{\cramped{n^2 \log n / \varepsilon}} $
  over the class $ \grcl{G} $.

Unfortunately, no local algorithm
  is able to implement the Metropolis rule
  over the class~$ \grcl{G} $:
  the rule is local only in the weak sense
  that an agent's next estimate~$ x_i(t) $
  depends on information present \emph{within distance~$ 2 $}
  of agent~$ i $ in round~$ t $,
  which is not local enough
  when the network is subject to change.

Indeed, since agent~$ j $ only knows
  its round~$ t $ degree $ \degree{j}{t} $
  at the end of round~$ t $,
  it has to wait until round~$ t + 1 $
  to share this information with its neighbors.
Any distributed implementation of this rule
  would require the communication links
  to evolve at a slow and regular pace.
As an example, we may assume that the links only change
  at rounds~$ t $ for which $ t \equiv r \bmod k $
  -- e.g., at even rounds.
Such conditions are all the more limitative
  in that they additionally require all agents
  to be loosely synchronized,
  as they have to agree on~$ k $
  and on the current round number
  -- at least modulo~$ k $.

The situation is even worse
  when the network is subject to unpredictable changes,
  as we need to warn all agents, ahead of time,
  about any upcoming topology change.
In effect, this amounts to having a global synchronizing signal
  precede every change in the communication topology.
For a topology changing in round~$ t_0 $,
  this differs little from starting an \emph{entirely new execution}
  with $ (x_1(t_0-1), \cdots, x_n(t_0-1)) $ for new input.

To paraphrase, provided the dynamic communication graph
  is sufficiently stable,
  one \enquote{can} implement the Metropolis rule over dynamic networks,
  but the execution is fully distributed
  only as long as no change occurs.

\section{Stabilizing weights for consensus}
\label{sec:mw}

Consider the update rule given by
\begin{equation}
  \label{eq:fw_onestep}
    x_i(t) = x_i(t-1)
    + \frac{1}{q_i} \sum_{j \in \Ns{i}{t}} (x_j(t-1) - x_i(t-1)) \enspace,
\end{equation}
  which we call the FixedWeight rule
  for the parameters $ q_1, \cdots, q_n > 0 $.
When $ \degree{i}{t} \leqslant q_i $,
  it acts as a sort of lazy version of the EqualNeighbor rule,
  where agent~$ i $ gives more importance
  to its own estimate $ x_i(t-1) $
  than to those of its neighbors.
Over the sub-class $ \grcl{C} \subset \grcl{G}_{\vert n} $
  of dynamic graphs for which
  $ \degree{i}{t} \leqslant q_i $
  holds for all agents at all times,
  the FixedWeight rule with parameters $ q_1 , \cdots , q_n $
  is shown in~\cite[Theorem 1.6]{chazelle2011sjco}
  to solve consensus with a convergence time in
  $ \cvtime[n]^{\grcl{C}}(\varepsilon)
  = \bigo{\sum_i q_i \cdot n \log n/\varepsilon} $.
In particular, using $ q_1 = q_2 = \cdots = q_n = n $
  yields a bound in
  $ \cramped{\bigo{n^3 \log n / \varepsilon}} $,
  comparable to the EqualNeighbor rule
  over \emph{static} networks.

Unfortunately, the ability for the agents to pick good values
  for the parameters $ q_1, \cdots, q_n$
  is limited by what they know, ahead of time,
  about the structure of the communication graph:
  parameters that are too small
  risk breaking the degree condition
  $ \degree{i}{t} \leqslant q_i $
  and cause the system to diverge,
  while parameters that are too large
  make convergence unacceptably slow
  if the network has a small degree.

Instead of relying on exogenous parameters,
  incompatible with a distributed approach,
  we propose making each agent~$ i $
  learn by itself what value of~$ q_i $
  work for the current execution.
We obtain the \defterm{MaxWeight} rule
  by replacing the fixed parameter $ q_i $
  with $ \dprime_i(t) \defeq
  \max \, \set{ \degree{i}{1}, \cdots, \degree{i}{t} } $,
  at each step of \cref{eq:fw_onestep}.

As each sequence $ \dprime_i(t) $ stabilizes
  over its limit $ \dprime_i \defeq \max_{t} \degree{i}{t} $,
  the MaxWeight rule eventually behaves
  like the FixedWeight rule with parameters
  $ \dprime_1, \cdots, \dprime_n $.
However, the MaxWeight rule
  can be implemented over the class $ \grcl{G} $
  with no additional assumption,
  resulting in the \defterm{MaxWeight} algorithm,
  given in \Cref{alg:mw}.

  \begin{algorithm}
    \SetKwData{Xi}{\ensuremath{\code{x}_{i}}}
    \SetKwData{Qi}{\ensuremath{\code{q}_{i}}}

    \KwIn{$ \mu_{i} \in \bbR $}

    \Initially{$ \Xi \gets \mu_{i} $ \; $ \Qi \gets 2 $ \;}

    \EachRound{%
      \Send $ m_i = \bothangles{\Xi} $ \;
      \Receive $ m_{j_1}, \ldots, m_{j_{d}} $
      \Comment*[r]{\( d \) neighbors}
      $ \Qi \gets \max (\Qi, d) $ \;
      $ \Xi \gets \Xi +
        \frac{1}{\Qi}
        \sum\limits_{k = 1}^d ( \code{x}_{j_k} - \code{x}_i )
      $ \;
      \Output $ \Xi $ \;
    }
    \caption{The MaxWeight algorithm, code for agent~$ i $}
    \label{alg:mw}
  \end{algorithm}

\begin{theorem}\label{theo:mw}
  The MaxWeight algorithm solves consensus
    over the class~$\grcl{G}$
    of dynamic graphs that are
    reflexive, bidirectional, and strongly connected in each round,
    with a convergence time of
    $\cvtime[n](\varepsilon) = \bigo{ n^4 \log n / \varepsilon } $
    for a system of $n$ agents.
\end{theorem}

\subsection{Technical preliminaries}

The proof of \Cref{theo:mw} will leverage
  some of the apparatus developed in~\cite{charron-bost2020ac},
  which we restate here briefly.

\subsubsection{Weighted geometries}

We fix a positive stochastic vector $ \vec{\pi} \in \bbR^{n} $,
  and we define the $\vec{\pi}$-\defterm{weighted Euclidian geometry}
  with the inner product
  $ \inner[\vec{\pi}]{}{} $
  and the norm $ \norm[\vec{\pi}]{} $:
  \begin{equation}
    \inner[\vec{\pi}]{\vec{u}}{\vec{v}} \defeq \sum_i \pi_i u_i v_i \enspace,
    \qquad
    \norm[\vec{\pi}]{\vec{v}}
    \defeq \sqrt{\inner[\vec{\pi}]{\vec{v}}{\vec{v}}} \enspace.
  \end{equation}
Controlling the dispersion of the estimates
  in an execution of the MaxWeight algorithm
  will specifically rely on the \defterm{variance}
  induced by this geometry:
  \begin{equation}
    \cramped{
    \var[\vec{\pi}] \vec{v} \defeq
    \norm[\vec{\pi}]{\vec{v}
    - \inner[\vec{\pi}]{\vec{v}}{\vec{1}} \vec{1}}^2
    = \norm[\vec{\pi}]{\vec{v}}^2 - \inner[\vec{\pi}]{\vec{v}}{\vec{1}}^{2}
    }
    \enspace.
  \end{equation}

We will need to switch between different
  geometries throughout an execution.
As a consequence, we will use the following lemma
  to relate variances with one another.

\begin{lemma}
  \label{lemma:variance-switch}
  Let $ \vec{\pi} $ and $ \vec{\pi}\safeprime $
    be two positive stochastic vectors of $ \bbR^n $.
  For any vector $ \vec{v} \in \bbR^{n} $,
    \begin{equation}\label{eq:different-variances}
      \var[\vec{\pi}\safeprime] \vec{v} \leqslant
      \max_i \frac{\pi\safeprime_i}{\pi_i} \var[\vec{\pi}] \vec{v}
      \enspace.
    \end{equation}
\end{lemma}
\begin{proof}
  For any vector $\vec{u}$ we have by definition
    \begin{equation}
      \norm[\vec{\pi}\safeprime]{\vec{u}}^2
      \leqslant \max_i \frac{\pi\safeprime_i}{\pi_i} \norm[\vec{\pi}]{\vec{u}}^2
      \enspace,
    \end{equation}
    and in particular for the vector
    $\vec{u} \defeq \vec{v} - \inner[\vec{\pi}]{\vec{v}}{\vec{1}}\vec{1}$.
  By the definition of the variance,
    we then have
    $\var[\vec{\pi}] \vec{v} = \norm[\vec{\pi}]{\vec{u}}^2$
    and $\var[\vec{\pi}\safeprime] \vec{v}
    = \var[\vec{\pi}\safeprime] \vec{u}
    \leqslant \norm[\vec{\pi}\safeprime]{\vec{u}}^2$,
    since adding a consensus vector
    does not change the variance,
    from which our claim follows.
\end{proof}

Moreover, we will have to relate
  the variance of a vector to its diameter,
  used to define the convergence time.
To this effect, we assume the next lemma,
  which generalizes \Cref{lemma:dispersion}
  to the $ \vec{\pi} $-geometries.

\begin{lemma}\label{lemma:weighted_dispersion}
  Let $ \vec{\pi} \in \bbR^{n} $
    be a positive stochastic vector.
  For any vector $ \vec{v} \in \bbR^{n} $,
  \begin{equation}
    2 \, \var[\vec{\pi}]{\vec{v}}
    < (\diam \vec{v})^2
    < \frac{4}{\min_i \pi_i} \var[\vec{\pi}]{\vec{v}}
    \enspace.
  \end{equation}
\end{lemma}

\subsubsection{Reversibility and Perron-Frobenius theory}

As a corollary of the celebrated Perron-Frobenius theorem,
  an irreducible matrix~$ \mat{A} $ admits
  a unique positive stochastic vector $ \vec{\pi}(\mat{A}) $
  satisfying $ \vec{\pi}(\mat{A}) = \mat{A}^{\transp} \vec{\pi}(\mat{A}) $,
  usually called the \defterm{Perron vector} of the matrix $ \mat{A} $.
We will say that a matrix is \defterm{reversible}
  if it is self-adjoint with respect to any
  $ \vec{\pi} $-weighted inner product,
  and in the case of an irreducible matrix
  the vector in question is necessarily its Perron vector.

By definition, a reversible matrix $ \mat{A} $
  is subject to the spectral theorem,
  and so is diagonalizable with real eigenvalues.
In the case of a stochastic matrix,
  $ \vec{1} $ is an eigenvector belonging to the eigenvalue $ 1 $,
  and the Rayleigh-Ritz variational characterization
  of the eigenvalues of the matrix $ \mat{A} $
  gives us the following lemma.

\begin{lemma}
  \label{lemma:weighted_var-eigenvalues}
  For an irreducible stochastic matrix $ \mat{A} $
    that is reversible and has positive diagonal entries,
    \begin{equation}\label{eq:weighted_var-eigenvalues}
      \forall \vec{v} \colon
      \var[\vec{\pi}(\mat{A})] \mat{A} \vec{v}
      \leqslant (1 - \spgap{\mat{A}})^2 \, \var[\vec{\pi}(\mat{A})] \vec{v}
      \enspace.
    \end{equation}
\end{lemma}

We will control the contraction of the estimates
  using \Cref{lemma:weighted_var-eigenvalues}
  with the following spectral bound,
  originally given in~\cite[Corollary 7]{charron-bost2020ac},
  which generalizes a previous bound from~\cite[Lemma 9]{nedic2009itac}.

\begin{lemma}
  \label{lemma:weighted_spectralgapbound}
  For an irreducible stochastic matrix $ \mat{A} \in \bbR^{n \times n} $
    that is reversible and has positive diagonal entries,
    \begin{equation}
      \spgap{\mat{A}} \geqslant \frac{\alpha(\mat{A})}{n-1} \enspace,
    \end{equation}
    where $ \alpha(\mat{A}) \defeq \min \,
    \set{\pi_i(\mat{A}) A_{ij} \where i,j \in [n] } \setminus \set{0} $.
\end{lemma}

\subsubsection{Proof of the theorem}

\begin{proof}
  We fix a dynamic graph $ \bbG \in \grcl{G}_{\vert n} $,
    and we let
    \begin{equation}
      \label{eq:dprime}
      \begin{gathered}
        \dprime_i(t) \defeq \max \,
        \set{ \degree{i}{\tau} \where 0 \leqslant \tau \leqslant t } \enspace,
        \qquad
        \dprime_i \defeq \max \set{ \degree{i}{t} \where t \in \bbN } \enspace,
        \\
        D\safeprime(t) \defeq \sum_{i = 1}^{n} \dprime_i(t) \enspace,
        \qquad
        D\safeprime \defeq \sum_{i = 1}^{n} \dprime_i \enspace, \\
        \Tl \defeq \set{ t \in \bbNz \where \exists i \in [n] :
        \dprime_i(t) \neq \dprime_i(t-1) } \enspace,
      \end{gathered}
    \end{equation}
    using the convention $ \degree{i}{0} = 2 $.
  The set $ \Tl $ has cardinal at most
    $ \abs{\Tl} \leqslant \sum_{i \in [n]} \dprime_i $,
    and so we let $ t_s \defeq \max \Tl $;
    for all $ t > t_s $,
    we have $ \dprime_i(t) = \dprime_i(t-1) $.

  Let us then pick arbitrary values
    $ \mu_1 , \cdots , \mu_n \in \bbR $,
    and consider the sequence of estimates
    $ \dvec{x} = \dvec{x}(\bbG,\vec{\mu}) $
    produced by the MaxWeight update rule
    for our chosen parameters:
    \begin{equation}
      \begin{gathered}
        x_i(0) = \mu_i \\
        x_i(t) = x_i ( t - 1 )
      + \frac{1}{\dprime_i(t)} \sum_{j \in \Ns{i}{t}}
      (x_j ( t - 1 ) - x_i ( t - 1 )) \enspace.
      \end{gathered}
    \end{equation}
  Since we have $ 2 \leqslant \dprime_i(t) \leqslant n $
    for all rounds $ t \in \bbNz $,
    the sequence $ \dvec{x} $ satisfies \Cref{prop:convergence}
    with uniform convexity parameter $ \alpha = 1/n $,
    and so it achieves asymptotic consensus
    within the convex hull of the set
    $ \set{ \mu_1 , \ldots , \mu_n } $.
  This shows that the MaxWeight rule,
    and thereby its implementing algorithm,
    solve the consensus problem over the class $ \grcl{G} $.

  It remains to control the convergence time.
  If we could bound the round~$t_s$
    -- say, by $t_s \leqslant f(n)$ for some function $f$ --
    then we could simply reuse the result from~\cite{chazelle2011sjco}
    about the FixedWeight update and deduce
    $\cvtime[n](\varepsilon) = \bigo{f(n) + n^3 \log n / \varepsilon}$.
  However, there clearly are
    some dynamic graphs in the class $\grcl{G}_{\vert n}$
    for which $t_s$ is arbitrary large,
    and we need to control the contraction of the estimates
    over the time window $ t \in [1 , \cdots , t_s] $.

  Let us then fix a disagreement threshold $ \varepsilon > 0 $,
    and consider the set of $ \varepsilon $-converged rounds:
    $ \mathcal{T}_{\varepsilon} \defeq
      \set{ t \in \bbN \where \diam \vec{x}(t)
      \leqslant \varepsilon \cdot \diam \vec{\mu} } $.
  The convexity of the sequence $ \dvec{x} $,
    along with the fact that it achieves asymptotic consensus,
    imply that this set is an unbounded interval of the integers,
    and we let
    $ t_{\varepsilon} \defeq \inf \mathcal{T}_{\varepsilon} $
    denote the earliest round at which
    the estimates agree with a relative error of~$ \varepsilon $.

  Let us denote by $ \mat{A}(t) $
    the round~$ t $ MaxWeight update matrix:
    \begin{equation}
      [\mat{A}(t)]_{ij} =
      \begin{cases}
        \frac{1}{\dprime_i(t)}
          &  i \neq j \in \Ns{i}{t} \\
        1 - \frac{\degree{i}{t}-1}{\dprime_i(t)}
          & j = i \\
        0 & j \notin \Ns{i}{t}
        \enspace.
      \end{cases}
    \end{equation}
  Its associated graph is $ G_{\mkern-2mu\mat{A}(t)} = \bbG(t) $,
    and by \Cref{assumption:graphs},
    this matrix is irreducible,
    with its Perron vector
    $ \vec{\pi}(t) \defeq \vec{\pi}(\mat{A}(t)) $
    given by
    \begin{equation*}
    \vec{\pi}(t)
    = ( \dprime_1 (t)/D'(t), \cdots, \dprime_n (t)/D'(t) )^{\transp}
    \enspace.
    \end{equation*}
  We can verify that the matrix $ \mat{A}(t) $
    is self-adjoint for the inner product
    $ \inner[\vec{\pi}(t)]{}{} $.
  As $ A_{ij}(t) \geqslant 1/\dprime_i(t) $
    for all positive entries of the matrix $ \mat{A}(t) $,
    \Cref{lemma:weighted_var-eigenvalues} gives us
    $ \gamma \defeq \inf_{t \in \bbNz}
      \spgap{\mat{A}(t)} \geqslant \cramped{1/(n-1)D\safeprime} $.

  Let us then define the potential
    $ \lyap( t ) \defeq \var[\vec{\pi}(t)]{\vec{x}(t)} $
    for positive $ t $.
  For a round~$t \neq 1$ outside of $\Tl$,
    the matrices $\mat{A}(t)$ and $\mat{A}(t-1)$ share their Perron vector,
    and with \Cref{lemma:var-eigenvalues} we have
    \begin{equation}\label{eq:mw_lyap_convex}
      \lyap( t ) \leqslant (1 - \gamma)^2 \lyap( t-1 )
      \enspace.
    \end{equation}

  \Cref{eq:mw_lyap_convex} provides some control
    over the dispersion of the estimates:
    for any temporal interval $[t, t\safeprime]$
    which does not intersect $\Tl$,
    applying \cref{eq:mw_lyap_convex} round-wise yields
    $ \lyap( t' ) \leqslant
      (1 - \gamma)^{2(t\safeprime - t)} \lyap( t ) $ --
    we find again that the rule achieves asymptotic consensus,
    as this $\lyap( t ) \! \to_t 0$.
  However, \cref{eq:mw_lyap_convex} alone
    cannot control the potential $\lyap( t )$
    across the entire execution.
  We use \Cref{lemma:variance-switch}
    to piece the variations of the potential over $ \Tl $:
    \begin{equation}
      \label{eq:mw:lyap}
      \forall t \geqslant 2:
      \lyap( t ) \leqslant
      \max_i \frac{\pi_i(t)}{\pi_i(t-1)} (1 - \gamma)^2 \lyap( t-1 )
      \enspace.
    \end{equation}

  The rounds for which $\vec{\pi}(t) \neq \vec{\pi}(t - 1)$
    are exactly those of $\Tl$,
    so \cref{eq:mw:lyap} is equivalent to \cref{eq:mw_lyap_convex}
    when $t \notin \Tl$.
  Applying \cref{eq:mw:lyap} over $\Tl$
    will induce a delay factor
    $ \beta_{\ell} \defeq
      \prod_{t \in \Tl}\max_i \frac{\pi_i(t)}{\pi_i(t-1)} $.
  Given $t \in \Tl$ we have

    \begin{align*}
      \max_i \frac{\pi_i(t)}{\pi_i(t-1)}
      &= \frac{\dprime(t-1)}{\dprime(t)}
      \max_i \frac{\dprime_i(t)}{\dprime_i(t-1)}  \\
      &\leqslant \frac{\dprime(t-1)}{\dprime(t)}
      \prod_i \frac{\dprime_i(t)}{\dprime_i(t-1)}
      \enspace,
    \end{align*}
    and with $\dprime_i(0) = 2$:
    \begin{equation}\label{eq:mw:beta}
      \beta_{\ell} \leqslant
      \frac{2 n}{\mathclap{\dprime}} \prod_i \frac{\dprime_i}{2}
      \enspace.
    \end{equation}

  Finally,
    \Cref{lemma:weighted_dispersion,lemma:weighted_var-eigenvalues} give us
    $\lyap( 1 ) \leqslant \frac{1}{2}
    \left((1 - \gamma) \, \diam \vec{\mu} \right)^2$.
  Together with \cref{eq:mw:lyap},
    we have for any $t \geqslant 2$:

  \begin{align*}
    \lyap( t ) &\leqslant
             \frac{1}{2}
             \prod_{\tau \leqslant t}
             \max_i \frac{\pi_i(\tau)}{\pi_i(\tau - 1)}
             \cdot
             \left( (1 - \gamma)^t \, \diam \vec{\mu} \right)^2 \\
             &\leqslant
             \frac{\beta_{\ell}}{2}
             \left( (1 - \gamma)^t \, \diam \vec{\mu} \right)^2 \\
             \intertext{and c using \Cref{lemma:dispersion},}
    \diam \vec{x}(t)
             &\leqslant
             \sqrt{\frac{2 \beta_{\ell}}{\min_i \pi_i(t)}}
             (1 - \gamma)^t \cdot \diam \vec{\mu}  \\
             \intertext{
               with \cref{eq:mw:beta}
               and the fact that $\pi_i(t) \geqslant 2/\dprime$,
             }
             &\leqslant
             \sqrt{n \prod_i \frac{\dprime_i}{2}}
             \, (1 - \gamma)^t \, \diam \vec{\mu}
             \enspace.
    \end{align*}

  We now let $\beta \defeq n \prod_i \frac{\dprime_i}{2}$.
  By definition of $t_{\varepsilon}$, we have
    \begin{gather*}
      \sqrt{\beta} \, (1 - \gamma)^{t_{\varepsilon}}
        \leqslant \varepsilon
        \enspace; \\
      \intertext{ equivalently, }
      t_{\varepsilon}
        \geqslant
        \frac{ \log (1 - \gamma) }{ \log (\varepsilon / \sqrt{\beta}) }
        \enspace,
    \end{gather*}
    and since $\log (1 - x) \leqslant -x$ when $x \in (0,1)$,
    \begin{equation}
      \label{eq:mw:almost-win}
      t_{\varepsilon}
      \leqslant \gamma^{-1} \log (\sqrt{\beta} / \varepsilon)
      \enspace.
    \end{equation}

  Inserting in this expression
    our bound over $ \gamma $,
    \begin{equation}
      \label{eq:mw:win}
      t_{\varepsilon}
      \leqslant \frac{(n-1) D'}{2}
      \left(
        \sum_i \log \dprime_i
        + \log n
        - 2 \log \varepsilon
        - n \log 2
      \right) \enspace,
    \end{equation}
    and with $\dprime_i \leqslant n$
    we have both $ D' \leqslant n^2 $ and
    $ \sum_i \log \dprime_i \leqslant n \log n $,
    which yields indeed
    $\cvtime[n](\varepsilon)
    = \bigo{n^4 \log n/\varepsilon}$.
\end{proof}

\section{The MaxMetropolis algorithm}
\label{sec:mm}

Girded with our stabilizing strategy,
  we now return to the problem of average consensus.
Recall that we obtained the Metropolis rule
  by applying the Metropolis-Hastings symmetrization
  to the EqualNeighbor update matrices.
Any consensus update rule can be given the same treatment;
  in particular, the symmetrized version of the FixedWeight rule
  -- where round~$ t $ neighbors $ i $ and $ j $
  apply the weight $ a_{ij}(t) = \tfrac{1}{\max(q_i,q_j)} $ --
  achieves asymptotic average consensus
  whenever the FixedWeight rule achieves asymptotic consensus.

This symmetrized FixedWeight rule is subject
  to the same limitations as the FixedWeight rule,
  which we now set out to circumvent
  as we did in the previous section.
However, in doing so we must also avoid
  the trappings of the Metropolis rule,
  where issues of information locality
  prevent the distributed implementation.
In particular, we observe that
  the symmetrized MaxWeight rule
  is no easier to implement than the Metropolis rule itself:
  for $ i $ and $ j $ neighbors in round~$ t $,
  the weight $ a_{ij}(t) $ depends on
  $ \dprime_j(t) = \max \, \set{ d_j(1), \cdots, d_j(t) } $;
  in particular, it depends on $ d_j(t) $
  like the Metropolis weight,
  which was the source of the problem
  with the Metropolis rule.

Our solution is to define the update rule
  in terms of an agent's largest degree
  \emph{up to the previous round},
  resulting in the \defterm{MaxMetropolis} update
  \begin{equation}
    \label{eq:maxmetropolis}
    x_i(t) = x_i(t-1)
           + \sum_{j \in \Ns{i}{t}}
             \frac{x_j(t-1) - x_i(t-1)}{\max(\dprime_i(t-1),\dprime_j(t-1))}
             \enspace,
  \end{equation}
  whose implementation by a local algorithm
  is given in \Cref{alg:mm}.

We can make a few immediate observations.
First, the MaxMetropolis rule
  defines symmetric update weights,
  and so the initial average
  is the only admissible consensus value.
Moreover, the weights are clearly stabilizing,
  and once they stop changing
  the MaxMetropolis rule behaves
  like the symmetrized FixedWeight rule
  with parameters $ \dprime_1 , \cdots , \dprime_n $.
From there, \Cref{prop:convergence} shows
  that \Cref{alg:mm} is an average consensus algorithm
  for the class $ \grcl{G} $.

However, %we now face the additional difficulty that
  the right-hand side of \cref{eq:maxmetropolis}
  no longer necessarily defines a \emph{convex} combination,
  as the self-weight $ a_{ii}(t) $ may be negative
  when $ \dprime_i(t) < \degree{i}{t} $.
In the worst case,
  the next estimate~$ x_i(t) $
  may leave the convex hull of the set
  $ \set{ x_1(t-1) , \cdots , x_n(t-1) } $,
  which delays the eventual convergence.
We show in \Cref{theo:mm},
  that this is only by at most a linear factor,
  when compared to a bound of
  $ \bigo{\cramped{n^3 \log n / \varepsilon}} $
  available for the symmetrized FixedWeight rule
  with parameters $ \dprime_1 , \cdots , \dprime_n $.

We note that the broken convexity does not fully explain
  the discrepancy of the convergence times
  between the Metropolis and MaxMetropolis rules,
  the former admitting a bound in~%
  $ \bigo{\cramped{n^2 \log n / \varepsilon} } $.
To explain the rest of the gap,
  observe that the Metropolis and MaxMetropolis rules
  take opposite approaches towards selecting weights.
The Metropolis update uses the exact degree of each agent,
  and is thus perfectly tailored to the graph in each round;
  on the other hand, the MaxMetropolis update
  only uses an \emph{upper bound} over the degree,
  a pessimistic approach that induces a slower convergence.

  \begin{algorithm}
    \SetKwData{Xi}{\ensuremath{\code{x}_{i}}}
    \SetKwData{Qi}{\ensuremath{\code{q}_{i}}}

    \KwIn{$ \mu_{i} \in \bbR $}

    \Initially{$ \Xi \gets \mu_{i} $ \; $ \Qi \gets 2 $ \;}

    \EachRound{%
      \Send $ m_i = \bothangles{\Xi,\Qi} $ \;
      \Receive $ m_{j_1}, \ldots, m_{j_{d}} $
      \Comment*[r]{\( d \) neighbors}
      $ \Xi \gets \Xi +
        \sum\limits_{k = 1}^d
        \frac{ \code{x}_{j_k} - \code{x}_i }{\max(\Qi,\code{q}_{j_k})}
      $ \;
      $ \Qi \gets \max (\Qi, d) $ \;
      \Output $ \Xi $ \;
    }
    \caption{The MaxMetropolis algorithm, code for agent~$ i $}
    \label{alg:mm}
  \end{algorithm}

\begin{theorem}\label{theo:mm}
  The MaxMetropolis algorithm solves average consensus
    over the class~$\grcl{G}$
    of dynamic graphs that are
    reflexive, bidirectional, and strongly connected in each round,
    with a convergence time of
    $\cvtime[n](\varepsilon) = \bigo{ n^4 \log n / \varepsilon } $
    for a system of $n$ agents.
\end{theorem}

In contrast with \Cref{theo:mw},
  the proof of \Cref{theo:mm}
  will only involve the usual geometry over the space $ \bbR^{n} $.
As a consequence,
  we briefly restate
  \Cref{lemma:dispersion,lemma:var-eigenvalues,lemma:spectralgapbound},
  specialized for the usual Euclidian norm $ \norm{} $.

\begin{lemma}\label{lemma:dispersion}
  For a non-null vector $ \vec{v} $
    for which $ \avgvec{\vec{v}} = 0 $, we have
  \begin{equation}
    \sqrt{2/n} \mkern2mu \norm{\vec{v}}
    < \diam \vec{v} < 2 \mkern2mu \norm{\vec{v}}
    \enspace.
  \end{equation}
\end{lemma}

\begin{lemma}
  \label{lemma:var-eigenvalues}
  For an irreducible stochastic matrix $ \mat{A} $
    that is symmetric and has positive diagonal entries,
    and any vector $ \vec{v} $
    for which $ \avgvec{\vec{v}} = 0 $,
    \begin{equation}
      \label{eq:var-eigenvalues}
      \norm{\mat{A} \vec{v}}
      \leqslant (1 - \spgap{\mat{A}}) \, \norm{\vec{v}}
      \enspace.
    \end{equation}
\end{lemma}

\begin{lemma}
  \label{lemma:spectralgapbound}
  For an irreducible stochastic matrix $ \mat{A} \in \bbR^{n \times n} $
    that is symmetric and has positive diagonal entries
    \begin{equation}
      \spgap{\mat{A}} \geqslant
      \frac{\phantom{A}\smash{A^{-}}}{n(n-1)}
      \enspace,
    \end{equation}
    where
    $ A^{-} \defeq \min \set{ A_{ij} \where i,j \in [n] } \setminus \set{0} $.
\end{lemma}

\begin{proof}[Proof of \Cref{theo:mm}]
  We fix a dynamic graph $ \bbG \in \grcl{G} $ of order~$ n $,
    and define $ \dprime_i(t) $, $ \dprime_i $, $ \Tl $, and $ t_s $
    as in \cref{eq:dprime},
    and recall from the proof of \Cref{theo:mw} that
    $ \abs{\Tl} \leqslant \sum_{i} \dprime_i $.

  Let us then fix an execution of \Cref{alg:mm}
    over the dynamic communication graph $ \bbG $,
    using input values $ \mu_1 , \cdots , \mu_n \in \bbR $.
  Without losing generality,
    we can assume $ \avgvec{\vec{\mu}} = 0 $,
    since a uniform translation of the input
    does not alter the relative positions of the estimates.

  An immediate induction reveals that the estimate vector
    satisfies the recurrence equation
    $ \vec{x}(t) = \mat{A}(t) \vec{x}(t-1) $,
    where the matrix $ \mat{A}(t) $
    is the round~$ t $ MaxMetropolis matrix
    for the dynamic graph $ \bbG $:
    \begin{equation}
      [\mat{A}(t)]_{ij} =
      \begin{cases}
        \frac{1}{\max(\dprime_i(t-1),\dprime_j(t-1))}
          &  i \neq j \in \Ns{i}{t} \\
        1 - \sum_{k = 1}^{n} \frac{1}{\max(\dprime_i(t-1),\dprime_k(t-1))}
          & j = i \\
        0 & j \notin \Ns{i}{t}
        \enspace.
      \end{cases}
    \end{equation}
  As such, it is a symmetric matrix
    for which $ \mat{A}(t) \vec{1} = \vec{1} $,
    and in paticular the average $ \avgvec{\vec{x}(t)} $
    is an invariant of the execution.
  Asymptotic consensus, if it happens,
    is necessarily over the average $ \avgvec{\vec{\mu}} = 0$.

  The matrix $ \mat{A}(t) $ results
    from a Metropolis-Hastings symmetrization,
    which implies
    $ A_{ii}(t) \geqslant 1 - \tfrac{\degree{i}{t} - 1}{\dprime_i(t-1)} $,
    and in particular for $ t \notin \Tl $
    we have $A_{ii}(t) \geqslant 1/n $.
  As the set $\Tl$ is finite,
    we let $t_s \defeq \max \Tl$,
    and the above holds in particular
    for all subsequent rounds $ t > t_s $.

  Let us then define another sequence $ \dvec{z} $,
    together with a dynamic graph $ \bbG' $,
    by $\vec{z}(k) \defeq \vec{x}(k + t_s + 1)$
    and $\bbG'(k) \defeq \bbG(k + t_s + 1)$,
    for each $ k \in \bbN $.
  The sequence $ \dvec{z} $
    satisfies the assumptions of \Cref{prop:convergence}
    for the dynamic graph $ \bbG' \in \grcl{G} $
    and uniform convexity parameter $ \alpha = 1/n $,
    and so it achieves asymptotic consensus
    and the sequence of estimates $ \dvec{x} $ does as well.
  Since the consensus value is necessarily the average
    $ \avgvec{ \vec{\mu} } $,
    we see that \Cref{alg:mm}
    is an average consensus algorithm
    for the class $ \grcl{G} $.

  To bound the convergence time,
    let us first remark that
    the diagonal entry~$ A_{ii}(t) $
    may be negative when
    $ \dprime_i(t) \neq \dprime_i(t-1) $,
    which can cause the estimate~$ x_i(t) $
    can then leave the convex hull of the set~%
    $ \set{ x_j(t-1) \where j \in \Ns{i}{t} } $.
  In fact, they can leave the convex hull
    of the set~$ \set{ x_j(t-1) \where j \in [n] } $,
    which moves the system away from consensus
    and delays the eventual convergence.

  To bound the total delay accrued in this manner,
    we fix a disagreement threshold $ \varepsilon > 0 $,
    and define the set
    $ \mathcal{T}_{\varepsilon} \defeq
      \set{ t \in \bbN \where \diam \vec{x}(t)
      \leqslant \varepsilon \cdot \diam \vec{\mu} } $.
  Since the system achieve asymptotic consensus,
    this set contains an unbounded interval,
    and we let
    $ t_{\varepsilon} \defeq
      \inf \, \set{ t \in \mathcal{T}_{\varepsilon} \where
      \forall \tau \geqslant t\colon \tau \in \mathcal{T}_{\varepsilon} } $.
  Remark that since the estimates can leave their convex hull,
    it is possible for $ t_{\varepsilon} $ to be greater than
    $ \inf \mathcal{T}_{\varepsilon} $.

  We then follow the variations of the quantity
    $ N(t) \defeq \norm{\vec{x}(t)} $ from one round to the next,
    distinguishing on whether $ t \in \Tl $ or not.
  When $ t \notin \Tl $,
    the update matrix $ \mat{A}(t) $ has positive diagonal entries,
    and by \Cref{lemma:spectralgapbound} we have
    $ \gamma \defeq \inf_{t \notin \Tl}
      \spgap{\mat{A}(t)} \geqslant \cramped{1/n^3} $.
  Using \Cref{lemma:var-eigenvalues}, we have
    \begin{equation}
      \label{eq:mm:lyap_convex}
      \forall t \notin \Tl \colon
      N( t ) \leqslant (1 - \gamma) N( t-1 )
      \enspace.
    \end{equation}

  For rounds $ t \in \Tl $, on the other hand,
    the update matrix $ \mat{A}(t) $
    may have negative entries,
    and we cannot use \Cref{lemma:var-eigenvalues}
    to control $ N(t) $.
  However, since the matrix $ \mat{A}(t) $ is symmetric,
    it is diagonalizable,
    and for any vector $ \vec{v} $, we have
    $ \norm{\mat{A}(t) \vec{v}}
      \leqslant \sprad{\mat{A}(t)} \norm{\vec{v}} $,
    which gives us
    \begin{equation}
      \label{eq:mm:radius_bound}
    \forall t \in \Tl \colon % TODO: fix in original
    N(t) \leqslant \sprad{\mat{A}(t)} N(t-1)
    \enspace.
    \end{equation}
  We note that \cref{eq:mm:radius_bound}
    holds in fact for all $ t \in \bbN $,
    but is strictly worse than \cref{eq:mm:lyap_convex}
    outside of $ \Tl $.

  To bound the spectral radius $ \sprad{\mat{A}(t)} $,
    we define $\nu_{t,i} \defeq 1 - \min(0, A_{ii}(t))$
    and $\nu_t \defeq \max_i \nu_{i,t}$.
  For any eigenvalue $ \lambda $ of the matrix $ \mat{A}(t) $,
    the quantity $(1 + \frac{\lambda - 1}{\nu_t})$
    is an eigenvalue of the stochastic matrix
    $\frac{1}{\nu_t}(\mat{A}(t) + (\nu_t - 1) \idmat)$,
    and so is less than $1$ in absolute value.
  We have $ 1 - 2 \, \nu_t \leqslant \lambda \leqslant 1 $,
    and so $\abs{\lambda} \leqslant 2 \, \nu_t - 1 \leqslant \nu_t^2$,
    the latter since $x^2 - 2 \, x + 1 \geqslant 0$ always holds.
  This holds for all eigenvalues of the matrix $ \mat{A}(t) $,
    and so we have
    \begin{equation}
      \label{eq:mm:lyap_learning}
      \forall t \in \Tl \colon
      N(t) \leqslant \nu_t^2 N(t-1)
      \enspace.
    \end{equation}

  The delay accrued during $ \Tl $
    will then depend on some factor
    $\beta_{\ell} \defeq \prod_{t \in \Tl} \nu_t^2$.
  To bound $ \nu_t $ for $ t \in \Tl $,
    we observe that
    $ \sum_{j \neq i} A_{ij}(t)
    \leqslant \frac{\degree{i}{t} - 1}{\dprime_i(t-1)}
    \leqslant \frac{\dprime_i(t)}{\dprime_i(t-1)} $
    for any $ i \in [n] $,
    which yields
    $ \nu_{i,t} \leqslant \frac{\dprime_i(t)}{\dprime_i(t-1)} $
    since $ \dprime_i(t) $ is weakly increasing.
  Given that $\nu_{i,t} \geqslant 1$, we have
    $\beta_{\ell}
    \leqslant \prod_{t \in \Tl} \prod_i \nu_{i,t}^2$,
    and since $\dprime_i(t) = \dprime_i(t-1)$ when $t \notin \Tl$,
    we finally have
    $ \beta_{\ell} \leqslant \left( \prod_i \tfrac{\dprime_i}{2} \right)^2 $

  Taking \cref{eq:mm:lyap_convex,eq:mm:lyap_learning} together, we have
    \begin{align*}
      N(t)
      &\leqslant
        \prod_{\mathclap{\tau \leqslant t : \tau \in \Tl}} \nu_{\tau}^2
        \prod_{\tau \leqslant t : \tau \notin \Tl} (1 - \spgap{\mat{A}(\tau)})
        \cdot N(0) \\
      &\leqslant
        \beta_{\ell} \,
        (1 - \gamma)^{t - \abs{\Tl}}
        \cdot N( 0 )
        \enspace.
     \end{align*}

  Using \Cref{lemma:dispersion},
    this gives us
    $ \diam \vec{x}(t) \leqslant
        2 \, \sqrt{n} \beta_{\ell} \,
        (1 - \gamma)^{t-\abs{\Tl}} \, \diam \vec{\mu}$.
  By definition of $ t_{\varepsilon} $,
    we have
    $ 2 \, \sqrt{n} \beta_{\ell} \,
      (1 - \gamma)^{t_{\varepsilon} - \abs{\Tl}} \leqslant \varepsilon $,
    from which we deduce that
    $ t_{\varepsilon} \leqslant
      \gamma^{-1}\log(2 \sqrt{n} \beta_{\ell} / \varepsilon) + \abs{\Tl} $.
  Using our upper bounds for
    $\abs{\Tl}$, $\gamma$, and $\beta_{\ell}$,
    \begin{equation}
      \label{eq:mm:win}
      t_{\varepsilon}
      \leqslant
      n^3
      \left(
        2 \sum_i \log \dprime_i
        - \log \varepsilon
        - (2 n - 1) \log 2
      \right)
      + \sum_i \dprime_i - 2n
      \enspace,
    \end{equation}
    and with $\dprime_i \leqslant n$
    the convergence time of the MaxMetropolis algorithm
    over the class $ \grcl{G} $ is in
    $\cvtime[n](\varepsilon) = \bigo{n^4 \log n/\varepsilon}$.
\end{proof}

\section{Discussion}
\label{sec:discussion}

In the preceding sections,
  we discussed a couple of well-studied update rules
  traditionally directed at the problems
  of consensus and average consensus
  in multi-agent networks with bidirectional interactions.
We argued that deploying these rules algorithmically
  is difficult over dynamic networks,
  where the communication links change often
  and in an unstructured manner.
The EqualNeighbor rule can take an exponentially long time
  before reaching consensus with a dynamic topology,
  whereas the Metropolis rule cannot even be implemented
  unless the fluctuations in the network
  are tightly choreographed,
  and not too frequent.

We introduced a parcimonious approach
  to cope with dynamic topologies,
  which requires no global control,
  no symmetry breaking devices,
  and no global information about the network,
  only making each agent keep track
  of its largest degree over time.
Defining the update rule in terms
  of the history of the network
  beyond its current configuration
  acts as a sort of low-pass filter,
  which dampens the effects of the fluctuations
  from pathological down to manageable.

Remarkably, adding \emph{a little} memory helps \emph{a lot}.
Out of all possible history-based strategies,
  we pick a rather crude one:
  each agent keeps track of a single parameter
  -- its historically largest degree, taking up to
  $\lceil \log n \rceil$ bits of memory --
  which it adjusts only upwards in the pursuit of correctness
  and never downwards in that of optimization.
Even so, it suffices to correct
  most of the deficiencies of the EqualNeighbor and Metropolis rules,
  at a moderate cost in the convergence time:
  compare the $ \bigo{\cramped{ n^3 \log n / \varepsilon }} $
  bound for EqualNeighbor over static networks
  and $ \bigo{\cramped{ n^2 \log n / \varepsilon }} $
  theoretical bound for Metropolis over dynamic networks
  with the $ \bigo{\cramped{ n^4 \log n / \varepsilon} } $ bounds
  for both the MaxWeight and MaxMetropolis algorithms.

We find bounds of the same order of magnitude
  for both algorithms, but, interestingly, for opposite reasons.
In an execution of the MaxWeight algorithm,
  the estimates keep moving towards a target value,
  but their convergence is set back
  when the target itself moves.
In contrast, the MaxMetropolis algorithm
  rigidly targets the average $ \avgvec{\vec{\mu}} $,
  and keeping the target constant
  sometimes requires individual estimates
  to move away from consensus,
  undoing earlier progress.

\small
\bibliography{ms}

\begin{thebibliography}{10}

\bibitem{alpern1995sjco}
Steve Alpern.
\newblock The {{Rendezvous Search Problem}}.
\newblock {\em SIAM Journal on Control and Optimization}, 33(3):673--683, 1995.
\newblock \href {https://doi.org/10.1137/S0363012993249195}
  {\path{doi:10.1137/S0363012993249195}}.

\bibitem{balch1998itra}
Tucker Balch and Ronald~C. Arkin.
\newblock Behavior-based formation control for multirobot teams.
\newblock {\em IEEE Transactions on Robotics and Automation}, 14(6):926--939,
  1998.
\newblock \href {https://doi.org/10.1109/70.736776}
  {\path{doi:10.1109/70.736776}}.

\bibitem{blondel2005p4icdc}
Vincent~D. Blondel, Julien~M. Hendrickx, Alex Olshevsky, and John~N.
  Tsitsiklis.
\newblock Convergence in {{Multiagent Coordination}}, {{Consensus}}, and
  {{Flocking}}.
\newblock In {\em Proceedings of the 44th {{IEEE Conference}} on {{Decision}}
  and {{Control}}}, pages 2996--3000, {Seville, Spain}, 2005. {IEEE}.
\newblock \href {https://doi.org/10.1109/CDC.2005.1582620}
  {\path{doi:10.1109/CDC.2005.1582620}}.

\bibitem{buck1938tqrob}
John~Bonner Buck.
\newblock Synchronous {{Rhythmic Flashing}} of {{Fireflies}}.
\newblock {\em The Quarterly Review of Biology}, 13(3):301--314, 1938.
\newblock \href {https://doi.org/10.1086/394562} {\path{doi:10.1086/394562}}.

\bibitem{cao2008sjco}
Ming Cao, A.~Stephen Morse, and Brian D.~O. Anderson.
\newblock Reaching a {{Consensus}} in a {{Dynamically Changing Environment}}:
  {{A Graphical Approach}}.
\newblock {\em SIAM Journal on Control and Optimization}, 47(2):575--600, 2008.
\newblock \href {https://doi.org/10.1137/060657005}
  {\path{doi:10.1137/060657005}}.

\bibitem{charalambous2016itcns}
Themistoklis Charalambous, Michael~G. Rabbat, Mikael Johansson, and
  Christoforos~N. Hadjicostis.
\newblock Distributed {{Finite}}-{{Time Computation}} of {{Digraph
  Parameters}}: {{Left}}-{{Eigenvector}}, {{Out}}-{{Degree}} and {{Spectrum}}.
\newblock {\em IEEE Transactions on Control of Network Systems}, 3(2):137--148,
  2016.
\newblock \href {https://doi.org/10.1109/TCNS.2015.2428411}
  {\path{doi:10.1109/TCNS.2015.2428411}}.

\bibitem{charron-bost2020ac}
Bernadette Charron-Bost.
\newblock Geometric {{Bounds}} for {{Convergence Rates}} of {{Averaging
  Algorithms}}.
\newblock {\em CoRR}, 2020.
\newblock \href {http://arxiv.org/abs/2007.04837} {\path{arXiv:2007.04837}}.

\bibitem{charron-bost2015alpi2}
Bernadette Charron-Bost, Matthias Függer, and Thomas Nowak.
\newblock Approximate {{Consensus}} in {{Highly Dynamic Networks}}: {{The
  Role}} of {{Averaging Algorithms}}.
\newblock In Magnús~M. Halldórsson, Kazuo Iwama, Naoki Kobayashi, and Bettina
  Speckmann, editors, {\em Automata, Languages, and Programming}, volume 9135
  of {\em Lecture Notes in Computer Science}, pages 528--539, {Berlin,
  Heidelberg}, 2015. {Springer Berlin Heidelberg}.
\newblock \href {https://doi.org/10.1007/978-3-662-47666-6_42}
  {\path{doi:10.1007/978-3-662-47666-6_42}}.

\bibitem{chatterjee1975p4sisiwp1}
Samprit Chatterjee.
\newblock Reaching a {{Consensus}}: {{Some Limit Theorems}}.
\newblock In {\em Proceedings of the 40th Session of the {{International
  Statistical Institute}}, {{Warsaw}}, {{Poland}}, 1975}, volume~3, pages
  156--160, {Warsaw, Poland}, 1975.

\bibitem{chatterjee1977jap}
Samprit Chatterjee and Eugene Seneta.
\newblock Towards {{Consensus}}: {{Some Convergence Theorems}} on {{Repeated
  Averaging}}.
\newblock {\em Journal of Applied Probability}, 14(1):89--97, 1977.
\newblock \href {https://doi.org/10.2307/3213262} {\path{doi:10.2307/3213262}}.

\bibitem{chazelle2011sjco}
Bernard Chazelle.
\newblock The {{Total}} s-{{Energy}} of a {{Multiagent System}}.
\newblock {\em SIAM Journal on Control and Optimization}, 49(4):1680--1706,
  2011.
\newblock \href {https://doi.org/10.1137/100791671}
  {\path{doi:10.1137/100791671}}.

\bibitem{cucker2007itac}
Felipe Cucker and Steve Smale.
\newblock Emergent {{Behavior}} in {{Flocks}}.
\newblock {\em IEEE Transactions on Automatic Control}, 52(5):852--862, 2007.
\newblock \href {https://doi.org/10.1109/TAC.2007.895842}
  {\path{doi:10.1109/TAC.2007.895842}}.

\bibitem{cybenko1989jopadc}
George Cybenko.
\newblock Dynamic load balancing for distributed memory multiprocessors.
\newblock {\em Journal of Parallel and Distributed Computing}, 7(2):279--301,
  1989.
\newblock \href {https://doi.org/10.1016/0743-7315(89)90021-X}
  {\path{doi:10.1016/0743-7315(89)90021-X}}.

\bibitem{daliot2003ss}
Ariel Daliot, Danny Dolev, and Hanna Parnas.
\newblock Self-{{Stabilizing Pulse Synchronization Inspired}} by {{Biological
  Pacemaker Networks}}.
\newblock In Shing-Tsaan Huang and Ted Herman, editors, {\em Self-{{Stabilizing
  Systems}}}, volume 2704 of {\em Lecture {{Notes}} in {{Computer Science}}},
  pages 32--48. {Springer Berlin Heidelberg}, {Berlin, Heidelberg}, 2003.
\newblock \href {https://doi.org/10.1007/3-540-45032-7_3}
  {\path{doi:10.1007/3-540-45032-7_3}}.

\bibitem{degroot1974jasa}
Morris~H. DeGroot.
\newblock Reaching a {{Consensus}}.
\newblock {\em Journal of the American Statistical Association},
  69(345):118--121, 1974.
\newblock \href {https://doi.org/10.2307/2285509} {\path{doi:10.2307/2285509}}.

\bibitem{dinitz2017ii2-iccc}
Michael Dinitz, Jeremy Fineman, Seth Gilbert, and Calvin Newport.
\newblock Load balancing with bounded convergence in dynamic networks.
\newblock In {\em {{IEEE INFOCOM}} 2017 - {{IEEE Conference}} on {{Computer
  Communications}}}, pages 1--9, {Atlanta, GA, USA}, 2017. {IEEE}.
\newblock \href {https://doi.org/10.1109/INFOCOM.2017.8057000}
  {\path{doi:10.1109/INFOCOM.2017.8057000}}.

\bibitem{hendrickx2006p1ismtns}
Julien~M. Hendrickx and Vincent~D. Blondel.
\newblock Convergence of {{Linear}} and {{Non}}-{{Linear Versions}} of
  {{Vicsek}}'s {{Model}}.
\newblock In {\em Proceedings of the 17th {{International Symposium}} on
  {{Mathematical Theory}} of {{Networks}} and {{Systems}}}, pages 1229--1240,
  {Kyoto (Japan)}, 2006.

\bibitem{jadbabaie2003itac}
Ali Jadbabaie, {Jie Lin}, and A.~Stephen Morse.
\newblock Coordination of groups of mobile autonomous agents using nearest
  neighbor rules.
\newblock {\em IEEE Transactions on Automatic Control}, 48(6):988--1001, 2003.
\newblock \href {https://doi.org/10.1109/TAC.2003.812781}
  {\path{doi:10.1109/TAC.2003.812781}}.

\bibitem{lamport1985ja}
Leslie Lamport and P.~Michael Melliar-Smith.
\newblock Synchronizing clocks in the presence of faults.
\newblock {\em Journal of the ACM (JACM)}, 32(1):52--78, 1985.
\newblock \href {https://doi.org/10.1145/2455.2457}
  {\path{doi:10.1145/2455.2457}}.

\bibitem{mirollo1990sjam}
Renato~E. Mirollo and Steven~H. Strogatz.
\newblock Synchronization of {{Pulse}}-{{Coupled Biological Oscillators}}.
\newblock {\em SIAM Journal on Applied Mathematics}, 50(6):1645--1662, 1990.
\newblock \href {https://doi.org/10.1137/0150098} {\path{doi:10.1137/0150098}}.

\bibitem{moreau2005itac}
Luc Moreau.
\newblock Stability of multiagent systems with time-dependent communication
  links.
\newblock {\em IEEE Transactions on Automatic Control}, 50(2):169--182, 2005.
\newblock \href {https://doi.org/10.1109/TAC.2004.841888}
  {\path{doi:10.1109/TAC.2004.841888}}.

\bibitem{nedic2017itac}
Angelia Nedic and Ji~Liu.
\newblock On {{Convergence Rate}} of {{Weighted}}-{{Averaging Dynamics}} for
  {{Consensus Problems}}.
\newblock {\em IEEE Transactions on Automatic Control}, 62(2):766--781,
  February 2017.
\newblock \href {https://doi.org/10.1109/TAC.2016.2572004}
  {\path{doi:10.1109/TAC.2016.2572004}}.

\bibitem{nedic2018pi}
Angelia Nedić, Alex Olshevsky, and Michael~G. Rabbat.
\newblock Network {{Topology}} and {{Communication}}-{{Computation Tradeoffs}}
  in {{Decentralized Optimization}}.
\newblock {\em Proceedings of the IEEE}, 106(5):953--976, 2018.
\newblock \href {https://doi.org/10.1109/JPROC.2018.2817461}
  {\path{doi:10.1109/JPROC.2018.2817461}}.

\bibitem{nedic2009itac}
Angelina Nedić, Alex Olshevsky, Asuman Ozdaglar, and John~N. Tsitsiklis.
\newblock On {{Distributed Averaging Algorithms}} and {{Quantization Effects}}.
\newblock {\em IEEE Transactions on Automatic Control}, 54(11):2506--2517,
  2009.
\newblock \href {https://doi.org/10.1109/TAC.2009.2031203}
  {\path{doi:10.1109/TAC.2009.2031203}}.

\bibitem{olfati-saber2004itac}
Reza Olfati-Saber and Richard~M. Murray.
\newblock Consensus {{Problems}} in {{Networks}} of {{Agents With Switching
  Topology}} and {{Time}}-{{Delays}}.
\newblock {\em IEEE Transactions on Automatic Control}, 49(9):1520--1533, 2004.
\newblock \href {https://doi.org/10.1109/TAC.2004.834113}
  {\path{doi:10.1109/TAC.2004.834113}}.

\bibitem{olshevsky2017sjco}
Alex Olshevsky.
\newblock Linear {{Time Average Consensus}} and {{Distributed Optimization}} on
  {{Fixed Graphs}}.
\newblock {\em SIAM Journal on Control and Optimization}, 55(6):3990--4014,
  2017.
\newblock \href {https://doi.org/10.1137/16M1076629}
  {\path{doi:10.1137/16M1076629}}.

\bibitem{olshevsky2011sr}
Alex Olshevsky and John~N. Tsitsiklis.
\newblock Convergence {{Speed}} in {{Distributed Consensus}} and {{Averaging}}.
\newblock {\em SIAM Review}, 53(4):747--772, 2011.
\newblock \href {https://doi.org/10.1137/110837462}
  {\path{doi:10.1137/110837462}}.

\bibitem{olshevsky2013itac}
Alex Olshevsky and John~N. Tsitsiklis.
\newblock Degree {{Fluctuations}} and the {{Convergence Time}} of {{Consensus
  Algorithms}}.
\newblock {\em IEEE Transactions on Automatic Control}, 58(10):2626--2631,
  2013.
\newblock \href {https://doi.org/10.1109/TAC.2013.2257969}
  {\path{doi:10.1109/TAC.2013.2257969}}.

\bibitem{qunli2006itc}
{Qun Li} and Daniela Rus.
\newblock Global clock synchronization in sensor networks.
\newblock {\em IEEE Transactions on Computers}, 55(2):214--226, 2006.
\newblock \href {https://doi.org/10.1109/TC.2006.25}
  {\path{doi:10.1109/TC.2006.25}}.

\bibitem{ren2007icta}
W.~Ren.
\newblock Consensus strategies for cooperative control of vehicle formations.
\newblock {\em IET Control Theory \& Applications}, 1(2):505--512, 2007.
\newblock \href {https://doi.org/10.1049/iet-cta:20050401}
  {\path{doi:10.1049/iet-cta:20050401}}.

\bibitem{reynolds1987cg}
Craig~W. Reynolds.
\newblock Flocks, {{Herds}}, and {{Schools}}: {{A Distributed Behavioral
  Model}}.
\newblock {\em Computer Graphics}, 21(4):10, 1987.

\bibitem{simeone2007jwcn}
Osvaldo Simeone and Umberto Spagnolini.
\newblock Distributed {{Time Synchronization}} in {{Wireless Sensor Networks}}
  with {{Coupled Discrete}}-{{Time Oscillators}}.
\newblock {\em EURASIP Journal on Wireless Communications and Networking},
  2007(1):057054, 2007.
\newblock \href {https://doi.org/10.1155/2007/57054}
  {\path{doi:10.1155/2007/57054}}.

\bibitem{sundaram20072acc}
Shreyas Sundaram and Christoforos~N. Hadjicostis.
\newblock Finite-{{Time Distributed Consensus}} in {{Graphs}} with
  {{Time}}-{{Invariant Topologies}}.
\newblock In {\em 2007 {{American Control Conference}}}, pages 711--716, 2007.
\newblock \href {https://doi.org/10.1109/ACC.2007.4282726}
  {\path{doi:10.1109/ACC.2007.4282726}}.

\bibitem{suzuki1999sjc}
Ichiro Suzuki and Masafumi Yamashita.
\newblock Distributed {{Anonymous Mobile Robots}}: {{Formation}} of {{Geometric
  Patterns}}.
\newblock {\em SIAM Journal on Computing}, 28(4):1347--1363, 1999.
\newblock \href {https://doi.org/10.1137/S009753979628292X}
  {\path{doi:10.1137/S009753979628292X}}.

\bibitem{tsitsiklis1984a}
John~N. Tsitsiklis.
\newblock {\em Problems in {{Decentralized Decision Making}} and
  {{Computation}}}.
\newblock PhD thesis, Massachusetts Institute of Technology, {Cambridge,
  Massachusetts, USA}, 1984.
\newblock URL: \url{https://www.mit.edu/~jnt/Papers/PhD-84-jnt.pdf}.

\bibitem{tsitsiklis1986itac}
John~N. Tsitsiklis, Dimitri~P. Bertsekas, and Michael Athans.
\newblock Distributed {{Asynchronous Deterministic}} and {{Stochastic Gradient
  Optimization Algorithms}}.
\newblock {\em IEEE Transactions on Automatic Control}, 31(9):803--812, 1986.
\newblock \href {https://doi.org/10.1109/TAC.1986.1104412}
  {\path{doi:10.1109/TAC.1986.1104412}}.

\bibitem{vicsek1995prl}
Tamás Vicsek, András Czirók, Eshel Ben-Jacob, Inon Cohen, and Ofer Shochet.
\newblock Novel {{Type}} of {{Phase Transition}} in a {{System}} of
  {{Self}}-{{Driven Particles}}.
\newblock {\em Physical Review Letters}, 75(6):1226--1229, 1995.
\newblock \href {https://doi.org/10.1103/PhysRevLett.75.1226}
  {\path{doi:10.1103/PhysRevLett.75.1226}}.

\bibitem{welch1988iac}
Jennifer~Lundelius Welch and Nancy Lynch.
\newblock A new fault-tolerant algorithm for clock synchronization.
\newblock {\em Information and Computation}, 77(1):1--36, 1988.
\newblock \href {https://doi.org/10.1016/0890-5401(88)90043-0}
  {\path{doi:10.1016/0890-5401(88)90043-0}}.

\bibitem{xiao2004s&cl}
Lin Xiao and Stephen Boyd.
\newblock Fast linear iterations for distributed averaging.
\newblock {\em Systems \& Control Letters}, 53(1):65--78, 2004.
\newblock \href {https://doi.org/10.1016/j.sysconle.2004.02.022}
  {\path{doi:10.1016/j.sysconle.2004.02.022}}.

\bibitem{xiao2005i2fisipsn2}
Lin Xiao, Stephen Boyd, and Sanjay Lall.
\newblock A {{Scheme}} for {{Robust Distributed Sensor Fusion Based}} on
  {{Average Consensus}}.
\newblock In {\em {{Fourth International Symposium}} on {{Information
  Processing}} in {{Sensor Networks}}, 2005.}, pages 63--70, {Los Angeles, CA,
  USA}, 2005. {IEEE}.
\newblock \href {https://doi.org/10.1109/IPSN.2005.1440896}
  {\path{doi:10.1109/IPSN.2005.1440896}}.

\end{thebibliography}

\end{document}